\documentstyle[epsf,12pt]{article}
\newlength{\dinwidth} 
\newlength{\dinmargin}
\def\3{\ss}
\begin{document}
\vspace{2cm}
\newcommand{\Gevsq}     {\mbox{${\rm GeV}^2$}}
\newcommand{\qsd}       {\mbox{${Q^2}$}}
\newcommand{\rhod}      {\mbox{${\rho^0}$}}
\newcommand{\wgp}       {\mbox{${W}$}}
\newcommand{\phid}      {\mbox{${\phi}$}}
\newcommand{\jpsid}     {\mbox{${J/\psi}$}}
\newcommand{\yjb}       {\mbox{${y_{_{JB}}}$}}
\newcommand{\ra}        {\mbox{$\rightarrow$}}
\newcommand{\gsp}       {\mbox{$\gamma^* p$}}
\newcommand{\xda}       {\mbox{$x_{_{DA}}$}}
\newcommand{\qda}       {\mbox{$Q^{2}_{_{DA}}$}}
\newcommand{\gsubh}     {\mbox{$\gamma_{_{H}}$}}
\newcommand{\thee}      {\mbox{$\theta_{e}$}}
\newcommand{\mrsdm}     {\mbox{MRSD$_-^{\prime}$\ }}
\newcommand{\mrsdz}     {\mbox{MRSD$_0^{\prime}$\ }}
\newcommand{\sleq} {\raisebox{-.6ex}{${\textstyle\stackrel{<}{\sim}}$}}
\newcommand{\sgeq} {\raisebox{-.6ex}{${\textstyle\stackrel{>}{\sim}}$}}
\renewcommand{\thefootnote}{\arabic{footnote}}
\def\pom{{\cal P}} 
\title{\vspace{2cm}
{\bf Measurement of the Reaction $\gamma^* p \rightarrow \phi p$
in Deep Inelastic $e^+p$ Scattering at HERA}
\\
\author{\rm ZEUS Collaboration \\}}
\date{}
\maketitle
\vspace{5 cm}
\begin{abstract}

The production of $\phi$ mesons in the reaction $e^{+}p \rightarrow e^{+}
\phi p$ ($\phi \rightarrow K^{+}K^{-}$), for $7 < Q^2 < 25$ GeV$^2$ and
for virtual photon-proton centre of mass energies ($W$) in the range
42-134 GeV, has been studied with the ZEUS detector at HERA.  When
compared to lower energy data at similar $Q^2$, the results show that the
$\gamma^*p \rightarrow \phid p$ cross section rises strongly with $W$.
This behaviour is similar to that previously found for the $\gamma^*p
\rightarrow \rho^0 p$ cross section. This strong dependence cannot be
explained by production through soft pomeron exchange. It is, however,
consistent with perturbative QCD expectations, where it reflects the rise
of the gluon momentum density in the proton at small $x$. The ratio of
$\sigma (\phi) / \sigma (\rho^0)$, which has previously been determined by
ZEUS to be 0.065 $\pm$ 0.013 (stat.) in photoproduction at a mean $W$ of
70 GeV, is measured to be 0.18 $\pm $ 0.05 (stat.) $\pm$ 0.03 (syst.)
at a mean $Q^2$ of 12.3 GeV$^2$ and mean $W$ of $\approx$
100 GeV and is thus approaching at large $Q^2$ the value of 2/9 predicted
from the quark charges of the vector mesons and a flavour independent
production mechanism.

\end{abstract}

\vspace{-20cm}
\begin{flushleft}
\tt DESY 96-067 \\
April 1996 \\
\end{flushleft}

\setcounter{page}{0}
\thispagestyle{empty}
\newpage

%

%===================================================================
%
%  MEMBER NAME  AUTH033 (ZEUS)     M  TEX
%
%  JH.: transformed to a format, which is suited as input for
%       CONVERT, which automatically creates author-indices
%
%  Don't remove lines starting with a percent sign %,
%  CONVERT may need them urgently !
%
%=====================================================================
\parindent0.cm
\def\3{\ss}
\footnotesize
\newcommand{\address}{ }
\renewcommand{\author}{ }
\pagenumbering{Roman}
                                    % this "%"s are for cosmetics only
                                                   %
\begin{center}
{                      \Large  The ZEUS Collaboration              }
\end{center}
  M.~Derrick,
  D.~Krakauer,
  S.~Magill,
  D.~Mikunas,
  B.~Musgrave,
  J.R.~Okrasinski,
  J.~Repond,
  R.~Stanek,
  R.L.~Talaga,
  H.~Zhang  \\
 {\it Argonne National Laboratory, Argonne, IL, USA}~$^{p}$
% ...
\newpage
% ...
$^{\    2}$ now at Univ. of Salerno and INFN Napoli, Italy \\
% ...
                                                           %
                                                           %
% \par         % if index listing & table fit to 1 page, put gap here
\newpage   % alternatively: go to newpage, if page is too small
                                                           %
% \institute_references_start    % do not touch or move this line !
                                                           %
\begin{tabular}[h]{rp{14cm}}
$^{a}$ &  supported by the Natural Sciences and Engineering Research
          Council of Canada (NSERC)  \\
$^{b}$ &  supported by the FCAR of Qu\'ebec, Canada  \\
$^{c}$ &  supported by the German Federal Ministry for Education and
          Science, Research and Technology (BMBF), under contract
          numbers 056BN19I, 056FR19P, 056HH19I, 056HH29I, 056SI79I \\
$^{d}$ &  supported by the MINERVA Gesellschaft f\"ur Forschung GmbH,
          the Israel Academy of Science and the U.S.-Israel Binational
          Science Foundation \\
$^{e}$ &  supported by the German Israeli Foundation, and
          by the Israel Academy of Science  \\
$^{f}$ &  supported by the Italian National Institute for Nuclear Physics
          (INFN) \\
$^{g}$ &  supported by the Japanese Ministry of Education, Science and
          Culture (the Monbusho) and its grants for Scientific Research \\
$^{h}$ &  supported by the Korean Ministry of Education and Korea Science
          and Engineering Foundation  \\
$^{i}$ &  supported by the Netherlands Foundation for Research on
          Matter (FOM) \\
$^{j}$ &  supported by the Polish State Committee for Scientific
          Research, grants No.~115/E-343/SPUB/P03/109/95, 2P03B 244
          08p02, p03, p04 and p05, and the Foundation for Polish-German
          Collaboration (proj. No. 506/92) \\
$^{k}$ &  supported by the Polish State Committee for Scientific
          Research (grant No. 2 P03B 083 08) and Foundation for
          Polish-German Collaboration  \\
$^{l}$ &  partially supported by the German Federal Ministry for
          Education and Science, Research and Technology (BMBF)  \\
$^{m}$ &  supported by the German Federal Ministry for Education and
          Science, Research and Technology (BMBF), and the Fund of
          Fundamental Research of Russian Ministry of Science and
          Education and by INTAS-Grant No. 93-63 \\
$^{n}$ &  supported by the Spanish Ministry of Education
          and Science through funds provided by CICYT \\
$^{o}$ &  supported by the Particle Physics and   
          Astronomy Research Council \\
$^{p}$ &  supported by the US Department of Energy \\
$^{q}$ &  supported by the US National Science Foundation \\
\end{tabular}
                                                           %
% \institute_references_end     % do not touch or move this line !

\newpage
\pagenumbering{arabic}
\setcounter{page}{1}
\normalsize

\section{\bf Introduction}                                                      
                                                                                
%\par
The elastic photoproduction of $\phi$ mesons, $\gamma p \rightarrow \phi p$,
has been studied 
in fixed target experiments~\cite{bauer,busen,egloff} and at HERA~\cite{zgp}
for photon-proton centre of mass (c.m.) energies ($\wgp$) up to 70 GeV.
For $\wgp > $ 10 GeV,
the reaction $\gamma p \rightarrow \phi p$ displays the characteristics
of a soft diffractive process: $s$-channel helicity conservation, 
a cross section rising weakly with $\wgp$ and an exponential $t$ dependence 
(where $t$ is the four-momentum transfer squared at the proton vertex)
with a slope $b(\wgp)$ which is also increasing slowly with $\wgp$.
Soft diffraction can be described by the exchange of a `soft'
pomeron Regge trajectory $\alpha(t) = \alpha(0)+\alpha^\prime t$ 
with an intercept $\alpha(0)=1.08$ and slope $\alpha^\prime=
0.25~{\rm GeV}^{-2}$.
The intercept is determined from fits~\cite{dl1}
to hadron-hadron total cross sections.
The same intercept also describes the energy dependence of the
photon-proton total cross section~\cite{sigtot}. In addition,
soft diffraction and the Vector Dominance Model can
describe the energy dependence of both $\phid$~\cite{zgp} and 
$\rhod$~\cite{rho} elastic photoproduction at HERA energies.

In contrast, the same soft pomeron fails to describe the 
recently measured energy dependences of the
cross sections at HERA for elastic $J/\psi$ 
photoproduction~\cite{psi}
and the exclusive production of $\rhod$ mesons~\cite{rhodis,h1rho} in
deep inelastic scattering (DIS) at
large values of $Q^2$, the negative of the four-momentum transfer squared 
of the exchanged virtual photon.
It also fails to describe the inclusive DIS diffractive cross 
section~\cite{kowalski}.
The rapid rise with energy of the cross sections for 
exclusive vector meson production is
consistent with recent perturbative QCD (pQCD)
calculations~\cite{ryskin,brod,forsh} in which the pomeron is treated as a 
perturbative two-gluon exchange. In such calculations the large scale may be 
the mass of the vector meson for $J/\psi$ photoproduction~\cite{ryskin}, the
$Q^2$ for exclusive DIS vector meson production~\cite{brod} 
or a large value of $t$~\cite{forsh}.

Brodsky et al.~\cite{brod} have studied the 
forward scattering cross section for DIS exclusive vector meson production
by applying pQCD 
in the double leading logarithm approximation (DLLA).
At high $Q^2$ and small Bjorken $x$ the 
vector mesons are expected to be produced dominantly by longitudinally 
polarised virtual photons with a
dependence for the longitudinal part of the differential cross section:
\begin{equation}
\label{brodsky}
 \left. \frac{d\sigma _L}{dt}\right|_{t=0}(\gamma^*N \rightarrow V^0 N) = 
\frac{A}{Q^6} \alpha_s^2(Q^2) \cdot \left| \left[ 1 + i\frac{\pi}{2} 
(\frac{d}{d~ln~x})
\right] xg(x,Q^2) \right|^2,
\end{equation}
where $A$ is a calculable constant 
and $xg(x, Q^2)$ is the momentum density of the gluon in the 
proton.  In view of the rapid rise of $xg(x)$ at small $x$, as derived from
HERA data \cite{f2data}, Eq. (1) predicts a rapid rise of the
cross section versus $W$ at fixed $Q^2$, substantially faster than the 
$W^{0.22}$ dependence expected for soft pomeron exchange.
The predicted $Q^2$ behaviour resulting from Eq. (1) is, however, weaker
than $Q^{-6}$ as the combined $Q^2$ dependences of the 
strong coupling constant and the gluon 
momentum density provide an additional factor of $\sim Q^{1-2}$ to the 
$Q^2$ dependence.

From the quark charges of the vector mesons and a flavour independent
production mechanism, the ratio  $\sigma(\phid) / \sigma (\rhod)$ 
of exclusive production cross sections
is expected to be 2/9~\cite{hjoos}. 
The pQCD prediction increases from 2/9 to 
2.4/9.0 at asymptotically large $Q^2$~\cite{brod,abram,frank}.
Experimentally, for photoproduction the ratio is found
to be 0.076 $\pm$ 0.010 at $\wgp$ = 17 GeV~\cite{egloff} and 0.065 $\pm$ 0.013
at 70 GeV~\cite{zgp}. 
At larger $Q^2$, NMC has  determined that $\sigma(\phid) / \sigma (\rhod)$
is $\approx$ 0.1 for $2 < 
Q^2 < 10$ GeV$^2$~\cite{nmc}. The NMC measurements are for $\wgp
\approx$ 15 GeV. It is of interest to determine this ratio at 
both large $Q^2$ and large $\wgp$.

This letter reports a measurement of the exclusive
cross section for $\phid$ mesons produced at large $Q^2$ 
by the process $\gamma^{*}p \rightarrow \phid p$ at HERA.
The data come from neutral current, deep inelastic
positron-proton scattering:
\begin{equation}                                                      
   e^+p \rightarrow e^+ \phid p
\end{equation} 
\noindent
in the $Q^2$ range 7 - 25 GeV$^2$, similar
to that of the earlier fixed target experiments \cite{nmc,emc}.
However, they cover a lower Bjorken $x$ 
region ($4\cdot10^{-4} < x < 1\cdot10^{-2}$) or, equivalently, a higher 
$\wgp$ region (42-134 GeV). The ratio of $\sigma(\phid) / \sigma (\rhod)$
is obtained by comparison to the ZEUS $\rhod$ measurement~\cite{rhodis}
at similar $Q^2$ values and the helicity decay distribution is studied.

%
%
%
%                                 SETUP
%
%
%
\section{Experimental setup}
\subsection{HERA}
During 1994 HERA operated with a proton beam energy $(E_p)$
of 820 GeV and a positron
beam energy ($E_e$) of 27.52 GeV. 
The positron and proton beams contained
153 colliding bunches
together with additional 17 proton and 15 positron unpaired
bunches. These additional bunches were
used for background studies. The time between bunch crossings
was 96 ns.
The typical instantaneous luminosity was $1.5 \cdot 10^{30}$
$\rm{cm^{-2}s^{-1}}$. The integrated luminosity for this study was 2.62
pb$^{-1}$, known to an accuracy of 0.08 pb$^{-1}$.
\subsection{The ZEUS detector}
A detailed description of the ZEUS detector can be found elsewhere~\cite{zstat}.
The main components used in this analysis are outlined below.

Charged particle momenta are reconstructed by the vertex 
detector (VXD)~\cite{vxd} and the central
tracking detector (CTD)~\cite{ctd}. These are  cylindrical drift chambers
placed in a magnetic field of 1.43 T
produced by a thin superconducting coil. The vertex detector surrounds
the beam pipe and consists of 120 radial cells, each with 12 sense wires.
The CTD surrounds the vertex detector and consists
of 72 cylindrical layers, organized in 9 superlayers
covering the polar angle\footnote{The ZEUS
coordinate system has positive-Z in the direction of flight of the protons and 
the X-axis is horizontal, pointing towards the center of HERA. The nominal 
interaction point is at X = Y = Z = 0.} 
 region $15^o < \theta < 164^o$.
Using the information from the CTD and the VXD
for the two-track events of this analysis,
the event vertex can be reconstructed with a
resolution of 0.4 cm in Z.
%and 0.14 cm in the transverse plane.
The transverse momentum resolution for tracks traversing all superlayers
is $\sigma(p_T)/p_T \simeq \sqrt{(0.005 p_T)^2 + (0.016)^2}$, with $p_T$
in GeV.

The high resolution uranium-scintillator calorimeter CAL~\cite{cal} is divided
into three parts, the forward (proton direction)
calorimeter (FCAL), the barrel calorimeter (BCAL)
and the rear (positron direction) calorimeter (RCAL), which 
cover polar angles from  $2.6^o$ to $36.7^o$,
$36.7^o$ to $129.1^o$, and $129.1^o$
to $176.2^o$, respectively. 
Each part consists of towers which are longitudinally subdivided
into electromagnetic (EMC) and hadronic (HAC) readout cells.
The transverse sizes are approximately $5\times20$ $\rm{cm^2}$ for the EMC cells
($10\times20$ $\rm{cm^2}$ in RCAL) and $20\times20$ $\rm{cm^2}$ for the HAC 
cells. From test beam data, energy resolutions of
$\sigma_{E}/E = 0.18/\sqrt{E}$ for electrons and $\sigma_{E}/E = 0.35/\sqrt{E}$
for hadrons have been obtained (with $E$ in GeV). In addition, the calorimeter 
cells provide time measurements with a time resolution below 1 ns for energy 
deposits greater than 4.5 GeV, a property used in background rejection.

The position of positrons scattered at small angles with respect to
the positron beam direction is
determined by the Small-angle Rear Tracking Detector (SRTD) which is attached
to the front face of the RCAL.
The SRTD
consists of two planes of scintillator strips, 1 cm wide and 0.5 cm thick, 
arranged in orthogonal directions and read out via optical fibres and 
photomultiplier tubes. It covers the region of 68$\times 68$ cm$^2$ in X 
and Y. A hole of 20$\times$20 cm$^2$ at the centre accommodates 
the beampipe. The SRTD 
is able to resolve clearly single minimum ionising particles and has 
a position resolution of 0.3 cm and a timing resolution
of better than 2 ns.

The luminosity was determined from the rate of the 
Bethe-Heitler process $e^{+} p \rightarrow e^{+} \gamma p$,
where the photon is measured by a calorimeter~\cite{lumi}
located at Z = $-104$ m in the HERA tunnel in the direction
of the positron beam. 

\subsection{Triggering}

Events were filtered online by a three level trigger 
system~\cite{zstat,wsmith}. 
At the first level, DIS events were selected by requiring a logical AND 
between two conditions based on energy deposits in the calorimeter. 
The first condition was the presence 
of an isolated electromagnetic energy deposit of greater than 
2.5~GeV.  The corresponding HAC energy was required to be either 
less than
0.95~GeV or no more than a third of the EMC energy.
The threshold values have been chosen
to give $>$99\% efficiency for detecting 
positrons with energy greater than 5 GeV
as determined by Monte Carlo studies.
The second condition required that the EMC section have an 
energy deposit greater 
than 3.75 GeV. Background from protons interacting outside the detector
was rejected using the time measurement of the energy deposits in the 
upstream veto counters and the SRTD.

At the second level trigger (SLT), 
background was further reduced using the measured 
times of energy deposits and the summed energies from the calorimeter. 
Events were accepted if
\begin{equation}
  \delta_{SLT} \equiv \sum_i E_i(1-\cos\theta_i) > 24
\:\:{\rm GeV} - 2E_{\gamma}
\end{equation}
where $E_i$ and $\theta_i$ are the energies and polar angles (with respect
to the nominal vertex position) of calorimeter cells, and $E_{\gamma}$
is the energy measured in the luminosity monitor photon calorimeter.
For perfect detector resolution and acceptance, 
$\delta_{SLT}$ is twice the positron beam energy (55~GeV) 
for DIS events while for photoproduction events,
where the scattered positron escapes down the beampipe,
$\delta_{SLT}$ peaks at much lower values.

The full event information 
was available at the third level trigger (TLT).
Tighter timing cuts as 
well as algorithms to remove beam-halo and cosmic muons were
applied.
The quantity $\delta_{TLT}$ was determined in the same manner as for
$\delta_{SLT}$ 
and was required to be $\delta_{TLT} > 25 \:\:{\rm GeV} - 2E_{\gamma}$.
Finally, DIS events were accepted if
a scattered positron candidate of energy greater than 4~GeV was found.

\section{Kinematics of exclusive $\phid$ production}

Figure 1 shows a schematic diagram for exclusive $\phid$ production in the 
reaction
\begin{equation}  
      e^+p \rightarrow e^+\phid N,
\end{equation} 
\noindent
where $N$ represents either a proton or a diffractively dissociated 
nucleonic system of mass $M_N$. 
The kinematics are described by the following variables:
  the negative of the squared four-momentum 
transfer carried by the virtual photon\footnote{In the $Q^2$ range
covered by this data sample, effects due to $Z^0$ exchange can be neglected.}
   $ Q^2=-q^2=-(k~-~k')^2$, where $k$ ($k^{\prime}$) is the 
four-momentum of the
incident (scattered) positron; the Bjorken variable $x =Q^2/2P\cdot q$,
where $P$ is the four-momentum of the incident proton;
the variable which describes the energy transfer to the hadronic             
final state  $   y =q\cdot P / k\cdot P$; 
the c.m. energy, $\sqrt{s}$, of the $ep$
system, where 
$ s = (k+P)^2 \approx 4E_e E_p = $ (300 GeV)$^2$; 
$W$, the c.m. energy  of the $\gamma^*p$ system:      
   $ W^2=(q+P)^2=Q^2(1-x)/x+M_p^2 \approx ys$,
where $M_p$ is the proton mass; and $ t^\prime=|t~-~t_{min}|$,
where $t$ is the four-momentum transfer squared, $t$ = 
$(P-P^{\prime})^2$, from the photon
 to the $\phid$,
$t_{min}$ is the maximum kinematically allowed value of $t$ and $P^{\prime}$ is
the four-momentum of the outgoing proton.
The squared transverse momentum $p_T^2$ of the 
$\phid$ with respect to the photon direction is 
a good approximation to $t^\prime$  since $t_{min}$ is
small, $|t_{min}| \ll 10^{-2}$~GeV$^2$.

In this analysis, the $\phid$ was observed in the decay $\phid \rightarrow 
K^+K^-$.  The three-momentum vector and energy ($E_{\phi}$)
of the $\phid$ was reconstructed from the
kaon three-momenta as determined from the tracking detectors and assuming
$K^{\pm}$ masses for the charged particles.
The production angles ($\theta_\phi$ and $\phi_\phi$) 
and momentum ($p_{\phi}$) of 
the $\phid$ and the angles of the scattered positron 
 (${\theta_e}^\prime$ and ${\phi_e}^\prime$), as determined with
the RCAL and SRTD, 
were used to reconstruct the kinematic variables $x, Q^2$, etc. Using 
energy and momentum conservation, the
energy of the scattered positron was determined 
from the relation
\begin{equation} 
   E_e^c = [2E_e - (E_{\phi} -p_{\phi}cos\theta_{\phi})]/
    (1 - cos\theta_e^\prime).
\end{equation}   
This relation assumes that $M_N=M_p$ and that the transverse momentum
of the proton is negligible compared to its longitudinal component.
The resulting resolution of the energy of the positron at a typical
energy of 26 GeV is less than $1\%$ compared to the one of the
direct measurement in the calorimeter of $5\%$.
The variable $y$ is calculated from the expression
$y$ = $ (E_{\phi} -p_{\phi}cos\theta_{\phi})/2E_e$.
The calculation of $p_T^2$ also uses the momentum of the $\phid$ and 
the corrected positron momentum:
$p_T^2$ = $(p_{ex}+p_{\phi x})^2 + (p_{ey}+p_{\phi y})^2$. 

%
%
%
%
%                           EVENT SELECTION
%
%
%
\section{Event selection}
The following off-line cuts were applied to select events from the reaction
$\gamma^* p\rightarrow \phi(\rightarrow K^+K^-) N$:
\begin{itemize}
\item select a scattered positron with an energy, as measured in the 
   calorimeter, greater than 5 GeV. 
   The positron identification algorithm is based on a neural network
   using information from the CAL and is described elsewhere~\cite{NN}.
   The efficiency of the identification algorithm is larger than 96\% for
   the final data sample; 

\item select events with a scattered positron whose impact point in the SRTD
      was outside the square of $24 \times 24~$cm$^2$   centered on the
      beam axis or events with an RCAL impact point outside the square of 
      $32 \times 32~$cm$^2$; this requirement
      controls the determination of the positron scattering angle; 
\item  require $\delta~=~\sum_i E_i(1 -\rm{cos}\theta_i) ~>~35$ GeV, where 
   the sum runs over all calorimeter cells; this cut reduces the
      radiative corrections;
\item require exactly two oppositely charged tracks associated 
      with a reconstructed vertex and not associated with the positron; 
\item require each track within the pseudorapidity\footnote{The pseudorapidity
      $\eta$ is defined as $\rm{\eta = -ln[tan(\frac{\theta}{2})]}$.}
      range $\left| \eta \right| < 1.75$ (corresponding to $20^o <\theta<160^o$)
      and with a transverse momentum greater than 150 MeV. These cuts select 
      the well understood and high efficiency region  of the tracking detector;
\item require that 
      the Z coordinate of the vertex is in the range $-50 $ to $ 40$ cm;
\item require 
      $E_{CAL}/P_{\phi}<$ 1.5, where $E_{CAL}$ is the calorimeter energy 
      excluding that due to the scattered positron 
      and $P_{\phi}$ is the sum of the absolute values of the momenta of the 
      two oppositely charged tracks. This cut suppresses backgrounds with
      additional calorimeter energy unmatched to the tracks;
\item reject photon conversion candidates ($\gamma \rightarrow e^+e^-$).
      The cut rejected two events in the $\phid$ mass range;
\item require  $p_T^2 < 0.6 \: \rm{GeV^2}$;
      this cut reduces the non-exclusive $\phid$ backgrounds. It also reduces
      background from proton dissociation which, from hadron-hadron diffractive 
      scattering, is expected to have a flatter $p_T^2$ distribution;
\item select $1.01 <M_{K^+K^-}<1.03$ GeV. 
      Figure~\ref{fig2}a shows a plot of the invariant mass of the 
      $K^+K^-$ system and demonstrates $\phid$
      production at the large values of $Q^2$ and $\wgp$
      of this experiment. 
\end{itemize}

Figure~\ref{fig2}b shows a scatter plot of $Q^2$ versus $x$ for the selected
events.
The acceptance at low $Q^2$ is limited by the requirement that the positron
is well contained in the detector: the selected events are therefore restricted 
to $Q^2>4~\Gevsq$.  The analysis presented here is limited to the 
region $7<Q^2<25~\Gevsq$. The track cuts limit the $y$ range to
$0.02 < y < 0.20$ ($42 < W < 134 \: \rm{GeV}$).
A total of 43 events passed all of these selection requirements. 
These are shown as the shaded histogram in Fig. 2a.
%
%
%
%               ACCEPTANCE. MC
%
%
%
\section{Monte Carlo simulation and acceptance calculation}

The reaction $e^+p \rightarrow e^+\phid p$ was modelled using the
Monte Carlo generator, DIPSI \cite{arneo}, which 
describes elastic vector meson
production in terms of pomeron exchange with the pomeron treated as
a colourless two-gluon system~\cite{ryskin}. The model 
assumes that the exchanged virtual 
photon fluctuates into a quark-antiquark pair
which then interacts with the two-gluon system.
The cross section is proportional to the square of the gluon 
momentum density in the proton. 
The generator DIPSI does not include radiative corrections.

The input vertex distribution was simulated in accordance
with that measured for unbiased photoproduction events.
The generated events were passed
through the ZEUS detector and trigger simulation programs
as well as through the analysis chain. 
The same offline cuts were used for the Monte Carlo events and for the data.
Good agreement is found between the Monte Carlo and data for the
distributions of the kinematic variables.

The simulated events were used to correct the data for
acceptance. The acceptance includes the geometric acceptance,
reconstruction efficiencies, detector efficiencies and resolution, 
corrections for the offline analysis cuts and a correction for the  
$M_{K^+K^-}$ cut. The acceptance
is shown in Fig.~\ref{fig2}c  as a function of $Q^2$;
in the region $7 < Q^2 < 25$ GeV$^2$, the acceptance
 varies between 44\% and 70\%. It drops sharply below  $Q^2=4~\Gevsq$ 
and it  also drops at small and large $y$. The 
acceptance is fairly constant at about 60\% as a function of $y$, $p_T^2$ or
$M_{K^+K^-}$ in the selected kinematic region.
The resolutions in the measured kinematic variables, as determined 
from the Monte Carlo events, are 
better than 4\% for $Q^2$ and 2\% for $y$, in
the $Q^2$, $y$ region of this analysis.

 The radiative corrections affecting the measured cross sections were 
calculated
analytically at various points in the $y-Q^2$ plane and were found to be
(10-15)\% for the selection cuts used in the analysis and for the
$Q^2$ and $\wgp$ dependences found in the data. 
They are taken into account in
the cross sections given below. The corrections were found to be the same for 
$\phid$ and $\rhod$ production to within 1\%.

\section{ Analysis, cross sections and results}
%
%
%
%                         PHI PHOTOPRODUCTION
%
%
%

\subsection{Background estimates}

Backgrounds to the exclusive reaction (2) are from $\phid$ events
with additional undetected particles, from $\rhod$ and $\omega$ production
and from proton dissociation events where the system $N$ in reaction
(4) has a small 
mass $M_N$ and does not deposit energy in the detector.
Studies of the unpaired bunches determined the beam-gas 
background to be negligible. The photoproduction background is also found to
be negligible because of the requirements on the $y$ range of the measurement
and the high energy for the scattered positron.

In order to estimate the non-resonant background,
a  non-relativistic Breit-Wigner (B-W), convoluted with a Gaussian, on 
a flat background is fit to the mass spectrum of Fig. 2a between 1.00 and 1.05 
GeV. The B-W width was fixed at the Particle Data Group (PDG) value of 4.43 
MeV~\cite{pdg}. The resulting $\phid$ mass is 1019.4 $\pm$ 0.4 MeV, 
to be compared with the PDG value of 1019.413 MeV.
The r.m.s. of the Gaussian
is found to be approximately $2$ MeV, consistent with the resolution
expected from tracking. The resulting
background under the $\phid$ signal (1.01 to 1.03 GeV) is estimated to
be ($14 \pm 7 ^{+4}_{-7}$)\%,
where the first number is the statistical and the second the systematic
uncertainty.
 The systematic error includes the uncertainties
from varying the shape of the background and the mass region that is fitted.

Since the proton was not detected, the proton dissociation background 
contribution had to be subtracted. 
Due to the limited statistics, the percentage of
proton dissociation background for the $\phid$ events
 was assumed to be the same as that determined for the $\rhod$ 
events, i.e. ($22 \pm 8 \pm 15$)\%~\cite{rhodis}. 

The overall factor to correct for background is then
$\Delta = (0.86 \times 0.78 = 0.67 \pm 0.17)$. Unless explicitly
stated otherwise, this background was subtracted as a constant
fraction for the cross sections given below.

\subsection{The $ep$ cross section}

The cross section, measured in the kinematic region defined above, is 
obtained from
\begin{equation}
 \sigma (e^+ p\rightarrow e^+ 
\phid p) = \frac{ \Delta }{C \cdot L_{int} \cdot B}
\sum_{i=1}^N \frac{1}{A_i},
\end{equation}
where $N$ (= 43) is the observed number of events after all cuts, 
$\Delta$ is the background correction factor,
$A_i$ is the bin-by-bin acceptance (which averages 58.7\% as discussed above),  
$L_{int}$ is the integrated luminosity of 2.62 pb$^{-1}$, $B$ = 0.491 is the
$\phid \rightarrow K^+K^-$ branching ratio~\cite{pdg} and 
$C$ = 1.12 is the average correction for QED radiative effects. 
The corrected $ep$ cross section for exclusive $\phid$ production at $\sqrt{s}$
= 300 GeV is 
\begin{displaymath}
   \sigma(e^+ p\rightarrow e^+ \phid p) = 
         0.034 \pm 0.007 ~(stat.) \pm 0.011~(syst.) \rm ~nb,
\end{displaymath}
integrated over the ranges $7 < Q^2 <$ 25 GeV$^2$,
$0.02 < y < 0.20$ and  $p_T^2 < 0.6$ GeV$^2$.

The quoted systematic uncertainty is derived from the following (the  
corresponding value is given in parentheses):
\begin{itemize}
\item the $E_{CAL}/P_{\phi}$ cuts used to remove non-exclusive backgrounds 
  were varied from 1.3 to 1.7. In an alternate background estimate, 
  tracks were matched to the calorimeter
  energy deposits and events containing an unmatched cluster with
  energy in excess of 0.3 or 0.4 GeV were discarded (5\%);
\item   
  the cut on the impact position of the positron was varied by 4 mm,
  the positron energy cut was varied and the $\delta$ cut was varied from
  30 to 40 GeV (13\%);
\item the cuts on the tracks were varied. The lower cut on the transverse 
  momentum was varied between 0.1 and 0.2 GeV and 
  different polar angle selections were made (4\%);
\item the cuts on the $M_{K^+K^-}$ region were varied by $\pm$ 2 MeV (7\%);
\item cuts were applied to the opening angle between the two charged 
 particles. Because of the low $Q$-value for the $\phid \rightarrow K^+K^-$ 
 decay, the charged decay particles have a small opening angle. This study
 checks the simulation of the ability of the CTD to resolve two close tracks 
 (5\%); 
%\item a cut on the helicity decay angle ($|$cos $\theta_H | <$ 0.8) was 
%   made (2\%);
\item the cut on the vertex position was varied by 10 cm (5\%);
\item 
the $Q^2$, $y$, $p_T^2$ and helicity decay dependence in the DIPSI Monte Carlo
 model was varied (3\%).
\end{itemize}
Adding these systematic uncertainties in quadrature (18\%) to those from the 
background correction (25\%), the luminosity determination and first level 
trigger efficiency (3.5\%) and the radiative corrections (10\%)
yields 32\% as the overall systematic uncertainty.

\subsection{The $\gamma^* p$ cross sections}

The $ep$
cross section can be converted to a $\gamma^*p$ cross section as follows.
The differential $ep$ cross section for one photon exchange can be expressed
in terms of the transverse and longitudinal virtual photoproduction 
cross sections as:
\begin{displaymath}
\frac{d^2\sigma (ep)}{dydQ^2} = \frac{\alpha}{2 \pi y Q^2} 
\left[\left( 1+(1-y)^2  \right) \cdot 
\sigma_T^{\gamma^* p}
(y, Q^2) + 2(1-y)\cdot \sigma_L^{\gamma^*p}(y, Q^2)\right].
\end{displaymath}
The virtual photon-proton cross section can then be written in terms of
the positron-proton differential cross section:
\begin{equation}
\label{sigtot}
\sigma (\gamma^*p \rightarrow \phid p) = 
(\sigma_T^{\gamma^*p}+\epsilon
\sigma_L^{\gamma^*p}) = \frac{1}{\Gamma_T} \frac{d^2\sigma (ep\rightarrow 
  e \phid p)}{dydQ^2},
\end{equation}
where $\Gamma_T$, the flux of transverse virtual photons, and $\epsilon$,
the ratio of the longitudinal to transverse virtual photon flux, are given by
\begin{displaymath}
\Gamma_T = \frac{\alpha \left( 1+(1-y)^2 \right) }
{2\pi y Q^2} ~~~{\rm and }~~~ 
 \epsilon = \frac{2(1-y)}{\left( 1+(1-y)^2 \right) } . 
\end{displaymath}
Throughout the kinematic range studied here, $\epsilon$ is in the range 
$0.97 < \epsilon < 1$.

Using Eq. (\ref{sigtot}), 
 $\sigma({\gamma^* p\rightarrow \phid p}$)
was determined with $\Gamma_T$ calculated from the $Q^2$, $x$  and $y$ values 
on an event-by-event basis. The resulting cross sections in two ranges of 
$Q^2$ are
\begin{displaymath}
   \sigma(\gamma^* p\rightarrow  \phid p) = 
10.3 \pm 2.2 ~(stat.) \rm ~nb~ for~ \it <Q \rm ^2> = 8.2~GeV^2~and~ \it <W> = 
\rm 94 ~GeV
\end{displaymath}
and
\begin{displaymath}
   \sigma(\gamma^* p\rightarrow  \phid p) = 
3.1 \pm 0.7 ~(stat.) \rm ~nb ~for~ \it <Q \rm ^2> = 14.7~ GeV^2~and ~ \it <W> = 
\rm 99~ GeV.
\end{displaymath}
\noindent
The 32\% overall systematic uncertainty
on $\sigma(ep)$ applies to both values for
 $\sigma({\gamma^* p\rightarrow \phid p}$). After correcting for the different
$<W>$ and assuming a $Q^2$ dependence of the form $Q^{-2 \alpha}$ one finds
$2 \alpha = 4.1 \pm 1.2 ({\rm stat.})$. This value
agrees, within errors, with the result found for the exclusive
$\rhod$ production $2 \alpha = 4.2 \pm 0.8 ({\rm stat.}) ^ {+1.4}_{-0.5}
{\rm (syst.)}$~\cite{rhodis}. 

\subsection{$\phid$ decay distribution } 

The $\phid$ s-channel helicity decay angular distribution,
$H$(cos$\theta_h,\phi_h,\Phi_h)$, can be used to 
determine the $\phid$ spin state \cite{guenter}, where
$\theta_h$ and $\phi_h$ are the polar and azimuthal angles, respectively, 
of the $K^+$ in the $\phid$ c.m. system and
$\Phi_h$ is the azimuthal angle of the $\phid$ production plane with
respect to the positron scattering plane. 
The quantisation axis is defined as the $\phid$ direction
in the $\gamma^*p$ c.m. system. After integrating over  $\phi_h$ and 
$\Phi_h$, the cos$\theta_h$ decay angular distribution,
shown in Fig. 2d, can be written as:
\begin{equation}                                                              
\label{helicity}
    \frac {1}{N}\frac{dN}{d\rm{cos} \it \theta_h} =
   \frac{3}{4}[1-r_{00}^{04}+(3 r_{00}^{04}-1)\rm{cos}^2\it \theta_{h}],
\end{equation}                
where $r_{00}^{04}$ is a particular linear combination of density matrix 
elements and represents
the probability that the $\phid$ 
is produced in the helicity zero state.

A maximum likelihood fit of the helicity cos$\theta_h$ distribution in
the range\footnote{Backgrounds due to photon conversions and $\rhod 
\rightarrow \pi^+\pi^-$ tend to populate the regions $|$cos$ \theta_h | > 
0.8$.}  $|$cos$ \theta_h| < 0.8$ to the form of Eq. (\ref{helicity}) yields
$r_{00}^{04}=0.76^{+0.11}_{-0.16}\pm 0.12 $
at $<Q^2>$ = 12.3 GeV$^2$ and $<\wgp>$ = 98 GeV.
This was not corrected for background, since
the dominant contribution is from $\phid$ production with
proton dissociation, which is expected to have a similar helicity.
The first uncertainty is statistical, and the second is derived 
from the variations of the result when different ranges in cos$\theta_h$
were used in the fit, when the systematic studies of section 6.2
were used or when a flat background of 15\% was included. 
This result is in sharp contrast to the measurement at $Q^2$ = 0~\cite{zgp},
where $r_{00}^{04}$ is compatible with zero,
 and indicates that the cross section for 
$\gamma ^* p  \rightarrow \phid p$ is dominated by 
$\phid$'s in the helicity zero state.
If $s$-channel helicity conservation (SCHC) is assumed, then $R = \sigma_L/
\sigma_T = r_{00}^{04}/\epsilon (1-r_{00}^{04})$. For DIS $\rhod$ production, 
$r_{00}^{04}=0.6 \pm 0.1^{+0.2}_{-0.1}$~\cite{rhodis}.
For the NMC $\phid$ data $r_{00}^{04}=0.84\pm0.18$ and $\epsilon$ = 
0.75~\cite{nmc}.

\subsection{The $\wgp$ dependence of the $\gamma^*p \rightarrow 
\phid p$ cross section}

Figure~\ref{fig3} shows a compilation~\cite{busen,nmc,lowe,lame,sand} of 
photoproduction and selected leptoproduction exclusive $\phid$ cross sections.
In this figure the cross section
\begin{equation}
 \sigma_T(\gsp \ra \phid p) + \sigma_L(\gsp \ra \phid p) = 
\frac {(1+R)}{(1+\epsilon R)} \cdot \sigma (\gsp \ra \phid p)
= (1-r_{00}^{04}+ \frac{r_{00}^{04}}{\epsilon})\cdot \sigma (\gsp \ra \phid p)
\end{equation} 
is plotted as a function of $W$. To obtain $R$, SCHC has been assumed, as
discussed above. In Fig.~\ref{fig3}, the cross sections at $Q^2$ = 0 are
shown assuming $R$ = 0. The results obtained in this 
analysis are shown at a mean flux-weighted $Q^2$ of 8.2 and 14.7 GeV$^2$, 
respectively. To compare with the NMC cross sections\footnote{Since the 
EMC~\cite{emc} and NMC
data cover approximately the same kinematic region, the more recent NMC 
data~\cite{nmc} have been chosen to make comparisons.} 
the NMC values were scaled to $Q^2$ = 8.2 (14.7) GeV$^2$ from $Q^2$ = 7.23
(11.35) GeV$^2$ using their measured $Q^{-4.54\pm0.78}$ dependence and 
$\epsilon$ = 0.75 was used. In addition, 
the NMC deuterium data have been corrected to obtain $\phid$ production off 
a nucleon using a factor\footnote{If instead, one corrects~\cite{sand} 
for the incoherent contribution (0.77$\pm$0.10) as 
well as a $d/p$ (normalised per nucleon) ratio of 0.77$\pm$ 0.21, as measured 
by E665~\cite{e665}, one obtains a factor of 1.0$\pm0.3$. These measurements
are for $\rhod$ production and
the $\phid$ data are assumed to behave in a similar manner.} 
of 0.94$\pm0.02$~\cite{frank}. 
To compare with results from the NMC experiment, which has determined
exclusive $\phid$ cross sections integrated over all $p_T^2$, 
a 4.5\% correction is applied to the ZEUS data 
to account for the cross section in the 
$p_T^2$ range between 0.6 and 1.0 GeV$^2$ based on the slope of the
distribution measured in the $\rhod$ analysis~\cite{rhodis}.

The $\gamma p \rightarrow \phid p$
cross section for 
real ($Q^2 = 0$) photons~\cite{zgp} shows only a slow rise, consistent
with  soft pomeron exchange
(as shown by the dashed line which represents a $W^{0.22}$ dependence).
At medium $Q^2 ~(< $ 8 GeV$^2$),
no high energy data are yet available.
At higher $Q^2$, the ZEUS values of the cross sections
are significantly larger than those of the NMC experiment.

Figure~\ref{fig3} shows that the cross sections rise strongly with increasing 
$\wgp$.  At $Q^2$ = 8.2 and 14.7 GeV$^2$,
the strong increase in the $\gamma^* p \rightarrow \phid p$ cross sections
between 12 (NMC data) and 100 GeV (ZEUS data) is in contrast to that 
expected from the Donnachie and Landshoff model~\cite{dl2} 
based on the energy dependence given by the soft pomeron.
However, it is similar to that observed previously for
the DIS $\rhod$ production~\cite{rhodis} and for J/$\psi$ 
photoproduction~\cite{psi}. It is also in qualitative agreement with pQCD 
calculations 
in which vector meson production is related to the square of the gluon 
momentum density~\cite{brod,frank}. As displayed in Eq. (1), these
calculations are for only the longitudinal part of the forward differential 
cross section.

A functional dependence $\sigma \sim W^k$ with $k=0.92 \pm 
0.08 \pm 0.16$ has recently been obtained for the inclusive $\gsp \ra X+N$ 
diffractive scattering cross sections for excitation masses $M_X$ up to 15 
GeV~\cite{kowalski}. The exclusive DIS $\phid$ vector meson data 
show a similar behaviour (see the solid line in Fig. \ref{fig3} which
is a $\sim W^{0.9}$ dependence).

\subsection{The ratio of $\phid$ to $\rhod$ production}

The ratio of the production cross sections for $\phid$ and $\rhod$ may be
determined using the previously published $\rhod$ measurement~\cite{rhodis}. 
Scaling the $\rhod$ data to the same $<Q^2>$ and $<W>$ with a $Q^{-4.2}$ and
$W^{0.8}$ dependence, and with the 
assumption that the proton dissociation background and the radiative 
corrections are identical, a value of $\sigma(\phid)/\sigma(\rhod)$ = 
0.18 $\pm$ 0.05 (stat.) $\pm$ 0.03 (syst.) is obtained 
at $<Q^2>$ = 12.3 GeV$^2$ and $<\wgp>$ = 98 GeV. The statistical error comes 
from adding in quadrature the statistical errors for the $\phid$ and the 
$\rhod$~\cite{rhodis} cross sections. The systematic error
comes from combining in quadrature the systematic errors 
for the $\phid$  and the $\rhod$~\cite{rhodis}. 
The correlated uncertainties which include contributions 
from the proton dissociation background and radiative corrections are excluded. 
This
result is shown in Fig. \ref{fig4} as a function of $Q^2$ along with similar 
data from the NMC experiment~\cite{nmc,sand}. 
Also shown is a recent determination at 
$\wgp \approx 70$ GeV and $Q^2 \approx 0$~\cite{zgp} giving 
$\sigma(\phid)/\sigma(\rhod)$ = 0.065$\pm$0.013 (stat.). 
As previously noted by the EMC 
experiment~\cite{emc}, the ratio for $Q^2$ values between 2 and 10 GeV$^2$ is 
larger than that at $Q^2 \approx 0$.
The measurement from ZEUS is consistent with the value of $2/9$ expected 
from the quark charges and a flavour independent production mechanism.
%for an SU(3) symmetric electromagnetic ccurrent
%and with the value predicted by pQCD.

\section{Summary } 

Exclusive $\phid$ production has been studied in deep inelastic 
$e^+p$ scattering at 
$Q^2$ values between 7 and 25 GeV$^2$ and in the
$\gamma^*p$ centre of mass 
energy ($W$) range from 42 to 134 GeV.
The $\gamma^* p \rightarrow \phid p$ cross sections at these large $Q^2$ values
are significantly larger than the NMC results, indicating a strong dependence
on  $W$ between the lower NMC energy and HERA, in contrast to the 
behaviour of the elastic $\phid$ photoproduction cross section.
 
The dependence is similar to that observed for
the reaction $\gamma^* p \rightarrow \rhod p$ and for the elastic
photoproduction of the J/$\psi$ and is in qualitative agreement with the 
strong energy dependence expected from pQCD 
calculations~\cite{ryskin,brod,abram} which relate these cross sections to 
the rise in the gluon momentum density in the proton at small $x$. 
A steep $W$ dependence is also observed in inclusive DIS
diffractive scattering~\cite{kowalski} and in the $\gsp$ total cross 
section\cite{f2data}.

The data suggest that the cross section for exclusive
vector meson production by real or virtual photons
has a strong energy dependence at large $W$-values whenever
a hard scale ($Q^2$ or $M^2_{J/\psi}$) is present.
The ratio of $\sigma(\phid) / 
\sigma(\rhod)$ of 0.18 $\pm$ 0.05 $\pm$ 0.03 at $<Q^2>$ of 12.3 GeV$^2$ 
is significantly larger
than the value observed in photoproduction at the same $W$ and is
approaching the expected  value of 2/9.

\section{Acknowledgements}
We thank the DESY Directorate for their strong support and
encouragement. The remarkable achievements of the HERA machine group
were essential for the successful completion of this work and are
gratefully appreciated.  We also acknowledge the many informative
discussions we have had with L. Frankfurt, P. Landshoff, A. Sandacz and 
M. Strikman.
                                                                               
%--------- REFERENCES -------------                                            
                 
\newpage
%
% fig.1
%
\newpage
\parskip 0mm
\begin{figure}
\epsfysize=18cm
\centerline{\epsffile{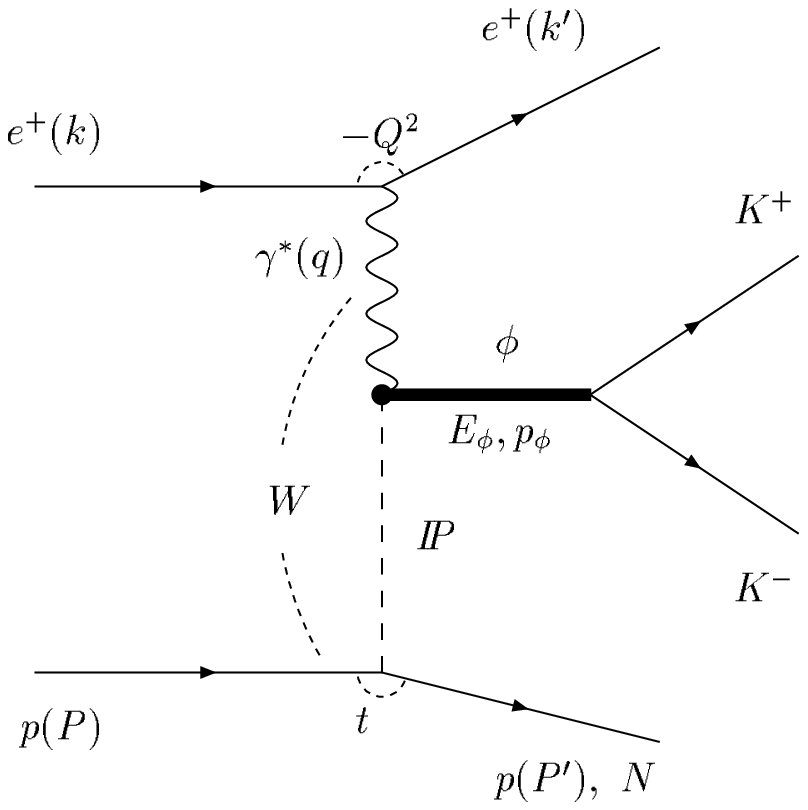}}
\caption{\label{fig1}
{Schematic diagram of exclusive $\phi$ production in deep inelastic 
$e^{+} p$ interactions.}
}
\end{figure}

\newpage
\parskip 0mm
\begin{figure}
\epsfysize=18cm
\centerline{\epsffile{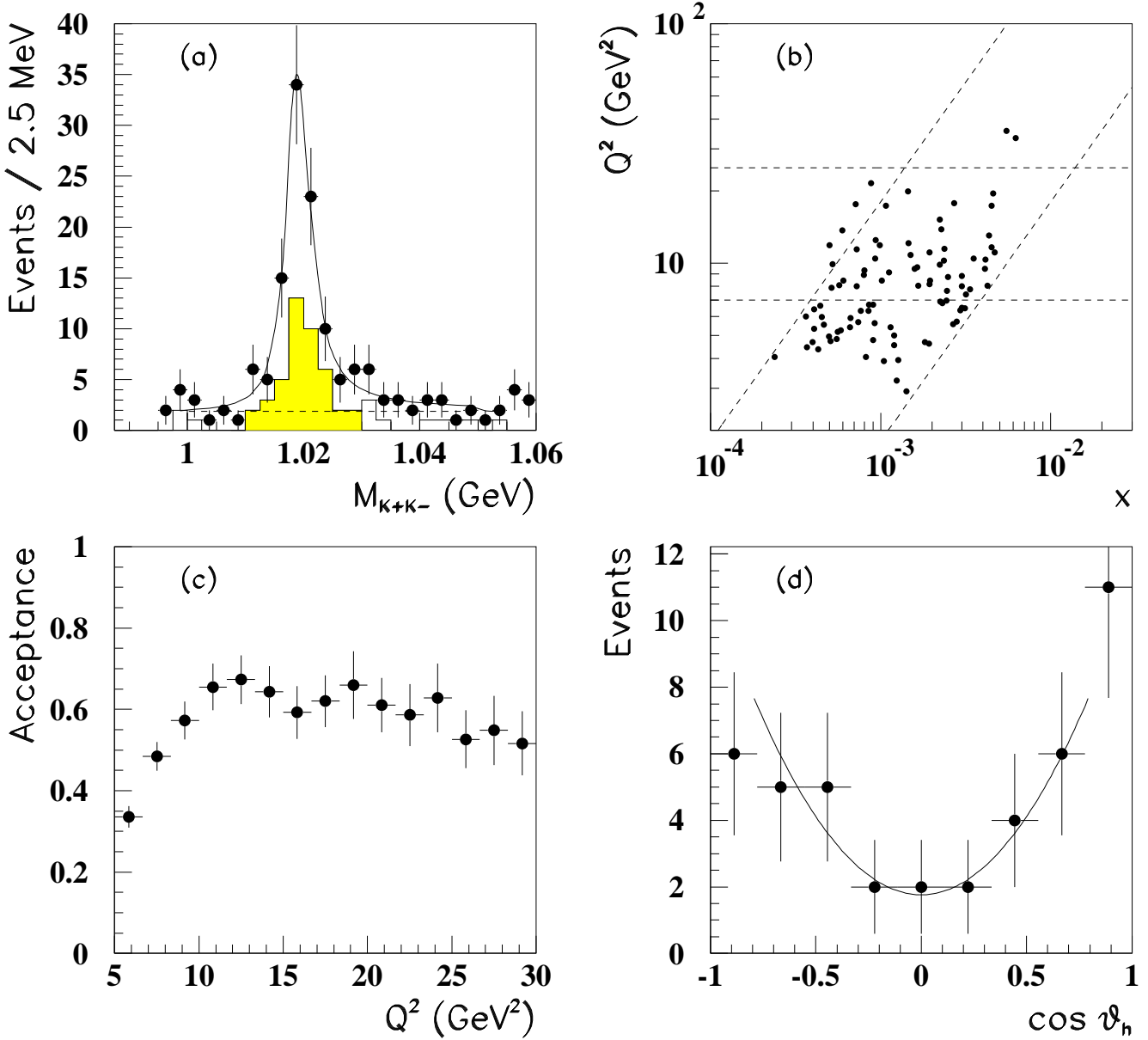}}
\caption{\label{fig2}{
(a) The invariant mass distribution for the $K^+K^-$ pairs; the curves are 
the best fit to
a Breit-Wigner convoluted with a Gaussian (solid line) over a flat background
(dashed line). The events in the $y-Q^2$ region of this analysis 
are shown as the hatched histogram;
(b) a scatter plot of $Q^2$ versus $x$ for the $\phid$ events.
The lines correspond to the region in 
$Q^2$ and $y$ selected for this analysis;
(c) the acceptance for DIS $\phid$ production as a function of $Q^2$ for 
events in the $y$ range of the cross section measurements; 
(d) the cos$\theta_h$ helicity angular distribution for candidate events. The 
acceptance in cos$\theta_h$ is flat over the region $|$cos$\theta_h |<0.8$.
The curve shows the result of the maximum likelihood fit for events in this 
range as described in the text.
}} 
\end{figure}

\newpage
\parskip 0mm
\begin{figure}
\epsfysize=18cm
\centerline{\epsffile{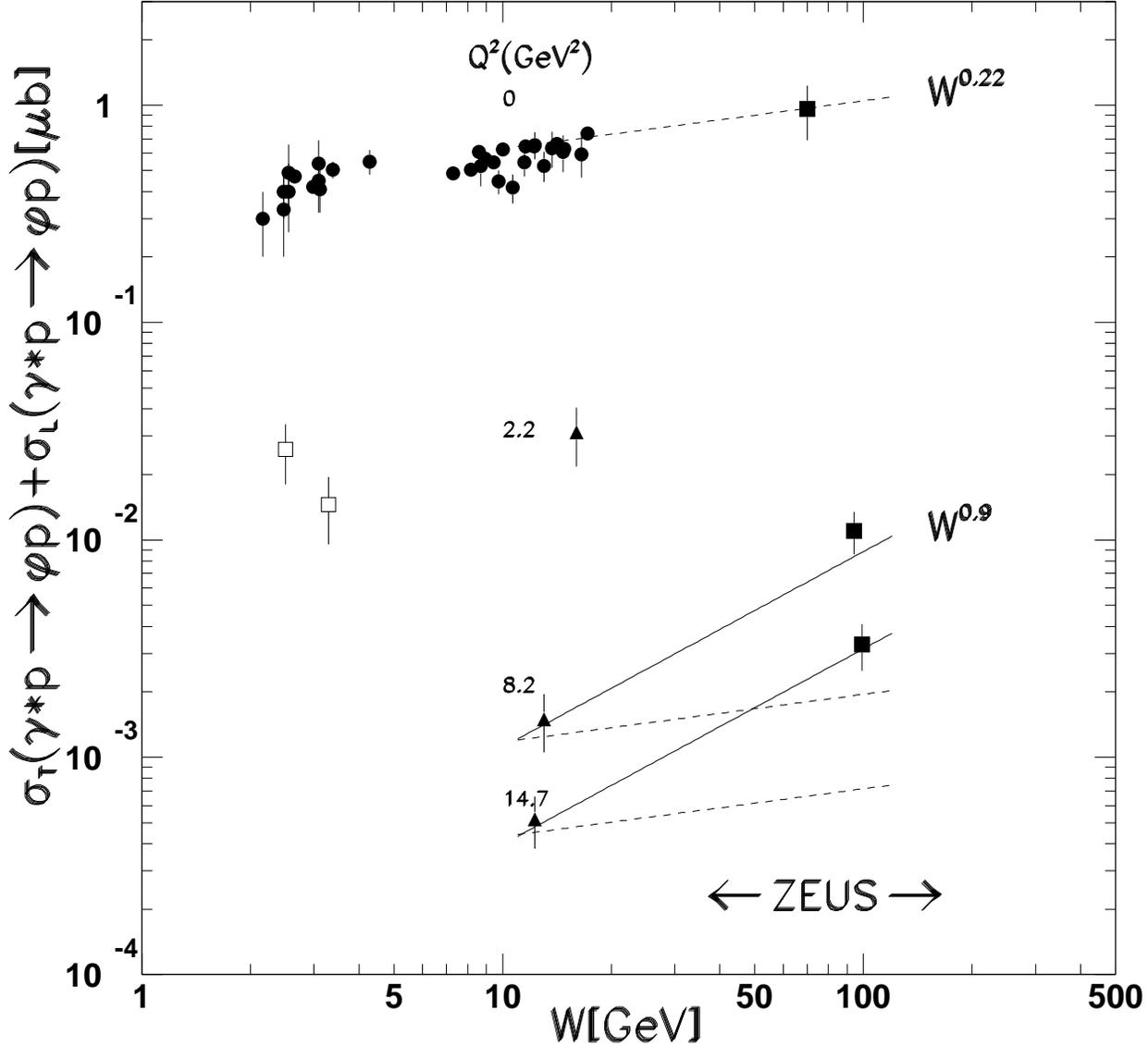}}
\caption{\label{fig3}{
The $\gamma^* p \rightarrow \phid p$ cross section ($\sigma_T + \sigma_L$) 
as a function of
$W$, the $\gamma^*p$ centre of mass energy, for several 
values of $Q^2$. The low energy data ($\wgp<$ 20 GeV, solid dots, open
squares and solid triangles) 
come from fixed target experiments [2,19,31-33]. 
The high energy data  ($\wgp>$ 50 GeV, solid squares) come from the 
ZEUS experiment [4] and the present analysis . The ZEUS data 
at $Q^2$ = 8.2 and 14.7 GeV$^2$ have an additional
correlated systematic uncertainty of 32\% (not shown); the data from Refs.
[19](solid triangles) and [32](open squares) have additional 20\% and 25\% 
normalisation uncertainties, respectively. The photoproduction data are 
shown assuming $R$ = 0. The NMC data [19] are scaled to the ZEUS $Q^2$ values
as described in the text. The dashed and solid lines are
drawn to guide the eye.
}} 
\end{figure}

\newpage
\parskip 0mm
\begin{figure}
\epsfysize=18cm
\centerline{\epsffile{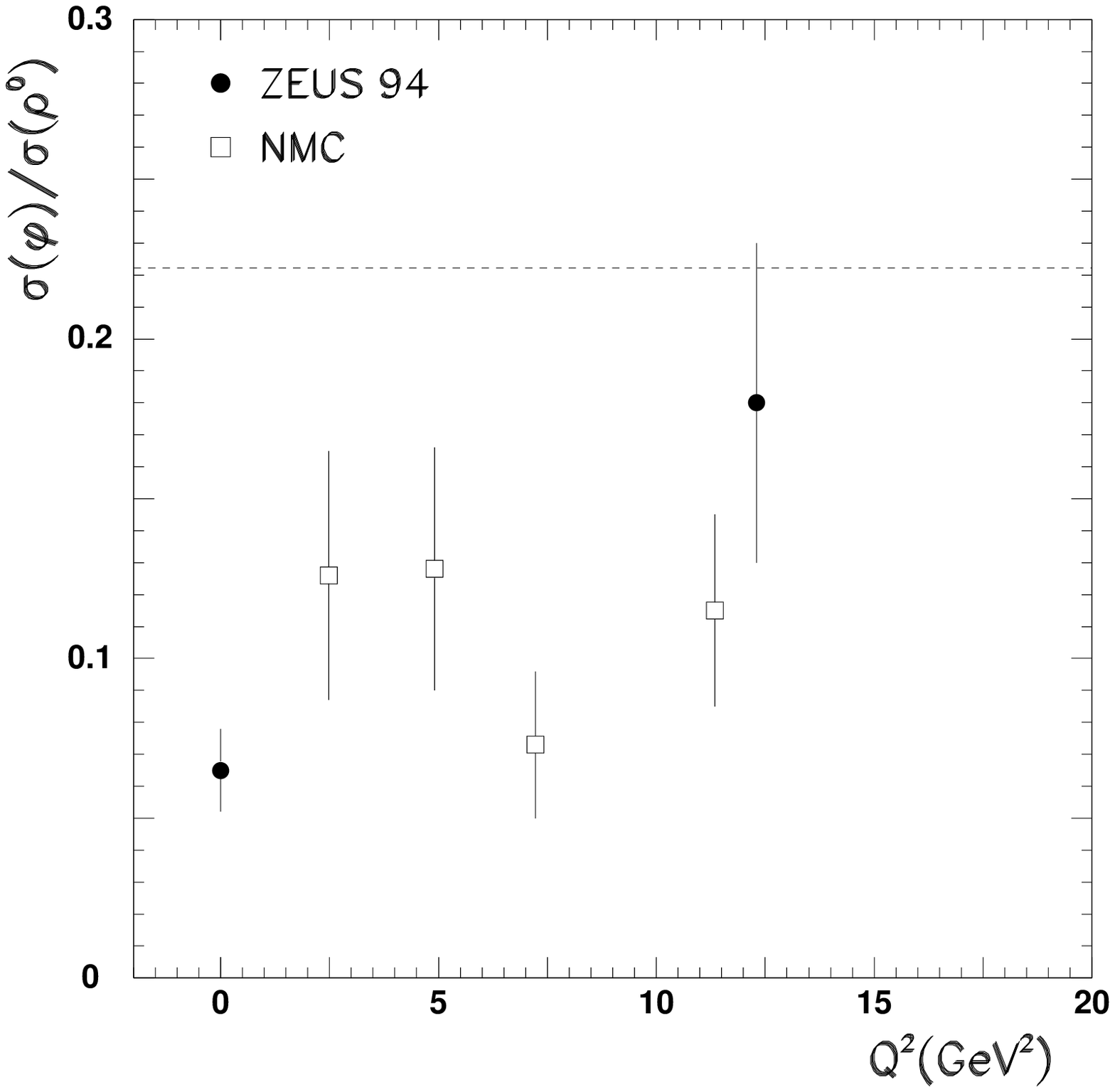}}
\caption{\label{fig4}{
The ratio of the cross section for exclusive $\phid$ to $\rhod$ production
is shown as a function of $Q^2$. The point at $Q^2$ =  12.3 GeV$^2$
is from the present analysis. Only the statistical error is shown. There is
an additional systematic error of 15\% arising from the different systematic 
errors for the $\phid$ and $\rhod$ analyses. 
The point at $Q^2 \approx $ 0 is also from this
experiment [4]. Also shown in the figure are data from the NMC 
collaboration [19]. Note that the ZEUS and NMC 
data points are at different $\gamma^* p$
c.m. energies ($W$), with the ZEUS data having 
$<W> \approx  100 $ GeV while
the NMC data points are at $<W> \approx $ 15 GeV. The horizontal
dashed line shows the value of 2/9 expected at high $Q^2$ (see text).
}} 
\end{figure}


\begin{thebibliography}{1}
\addcontentsline{toc}{chapter}{Bibliografy}
%%%%%
\bibitem{bauer} T.H. Bauer et al., Rev. Mod. Phys. 50 (1978) 261.
\bibitem{busen} J. Busenitz et al., Phys. Rev. D40 (1989) 1.
\bibitem{egloff} R.M. Egloff et al., Phys. Rev. Lett. 43 (1979) 657.
\bibitem{zgp} ZEUS Collab., M. Derrick et al., DESY 96-002, 
  to be published in Phys. Lett. B (1996).
\bibitem{dl1} A. Donnachie and P.V. Landshoff, Nucl. Phys. B244 (1984) 322;
   \newline Phys. Lett. B296 (1992) 227.
\bibitem{sigtot} ZEUS Collab., M. Derrick et al., Z. Phys. C63 (1994) 391;
   \newline H1 Collab., S. Aid et al., Z. Phys. C69 (1995) 27.
\bibitem{rho} ZEUS Collab., M. Derrick et al., Z. Phys. C69 (1995) 39;
   \newline H1 Collab., S. Aid et al., DESY 95-251.
\bibitem{psi} H1 Collab., T. Ahmed et al., Phys. Lett. B338 (1994) 507; 
    \newline ZEUS Collab., M. Derrick et al., Phys. Lett. B350 (1995) 120.
\bibitem{rhodis} ZEUS Collab., M. Derrick et al., Phys. Lett. B356 (1995) 601.
\bibitem{h1rho} H1 Collab., S. Aid, DESY 96-023 (1996).
\bibitem{kowalski} ZEUS Collab., M. Derrick et al., DESY 96-018,
    to be published in Z. Phys. C (1996).
\bibitem{ryskin} M. G. Ryskin, Z. Phys. C57 (1993) 89.
\bibitem{brod} S. J. Brodsky et al., Phys. Rev. D50 (1994) 3134.
\bibitem{forsh} J. R. Forshaw and M. G. Ryskin, Z. Phys. C68 (1995) 137;
   \newline ~J. Bartels et al, DESY 95-253 (1995).
\bibitem{f2data} ZEUS Collab., M. Derrick et al., Z. Phys. C65 (1995) 379;
    \newline  H1 Collab., T. Ahmed et al.,   
                           Nucl. Phys. B439 (1995) 471.
\bibitem{hjoos} H. Joos, Acta Physica Austriaca, Suppl. IV (1967) 320.
\bibitem{abram} H. Abramowicz et al., DESY 95-047 (1995); 
    M. Strikman, private communication.
\bibitem{frank} L. Frankfurt, W. Koepf and M. Strikman, TAUP-2290-95, 
 hep-ph/9509311 (1995).
\bibitem{nmc} NMC Collab., M. Arneodo et al., Nucl. Phys. B429 (1994) 503.
\bibitem{emc} EMC Collab., J. Ashman et al., Z. Phys. C39 (1988) 169. 
\bibitem{zstat} ZEUS Collab., The ZEUS Detector Status Report 1993;
  \newline ZEUS Collab., M. Derrick et al., DESY 95-193 (1995).
\bibitem{vxd} C. Alvisi et al., Nucl. Instr. Meth. A305 (1991) 30. 
\bibitem{ctd} C. B. Brooks et al., Nucl. Instr. Meth. A283 (1989)
  477;  \newline B. Foster et al., Nucl. Instr. Meth. A338 (1994) 254. 
\bibitem{cal} M. Derrick et al., Nucl. Instr. Meth. A309 (1991) 77;\\
  A. Andresen et. al., Nucl. Instr. Meth. A309 (1991) 101;\\
  A. Bernstein et. al., Nucl. Instr. Meth. A336 (1993) 23.
\bibitem{lumi} D. Kisielewska et al., DESY-HERA 85-25 (1985);\\
  J.Andruszk\'ow et al., DESY 92-066 (1992).
\bibitem{wsmith} W. H. Smith et al., Nucl. Instr. Meth. A355 (1995) 278.
\bibitem{NN} H. Abramowicz, A. Caldwell and R. Sinkus, Nucl. Instr. Meth. 
   A365 (1995) 508;
   \newline R. Sinkus, Ph. D. Thesis, University of Hamburg, 1996.
\bibitem{arneo}  M. Arneodo, L. Lamberti and M. G. Ryskin, 
            submitted to Comp. Phys. Comm. (1995). 
\bibitem{pdg} Review of Particle Properties, Particle Data Group, 
  L. Montanet et al., Phys. Rev.    D50 (1994) 1173.
\bibitem{guenter} K. Schilling, P. Seyboth and G. Wolf, Nucl. Phys. B15 
(1970) 397; \newline K. Schilling and  G. Wolf, Nucl. Phys. B61 (1973) 381.
\bibitem{lowe} S.I. Alekhin et al., CERN-HERA 87-01 (1987); \\
   ``Total Cross-Sections for Reactions of High Energy Particles'',
  Landolt-B\"ornstein, New Series, Vol. 12b, editor H. Schopper (1987).
\bibitem{lame} D. G. Cassel et al., Phys. Rev. D24 (1981) 2787.
\bibitem{sand}   A. Sandacz, private communication.
\bibitem{e665} Fermilab E665 Collab., M. R. Adams et al., Phys. Rev. Lett. 74 
(1995) 1525.
\bibitem{dl2} A. Donnachie and P. V. Landshoff, Nucl. Phys. B311 (1989) 509;
    \newline    Phys. Lett. B185 (1987) 403; 
    \newline    Phys. Lett. B348 (1995) 213.
                                                  
\end{thebibliography}
\end{document}